# Modulation of calcium signaling on demand to decipher the molecular mechanisms of primary aldosteronism

Bakhta Fedlaoui[1], Teresa Cosentino[1], Zeina R. Al Sayed[1], Rita Alexandre Coelho[2], Isabelle Giscos-Douriez[1], Nicolo Faedda[1], May Fayad[1], Jean-Sebastien Hulot[1,3], Chris Magnus[4], Scott M Sternson[4], Simon Travers-Allard[1,5], Stephanie Baron[5], David Penton[2], Fabio L Fernandes-Rosa[1], Maria-Christina Zennaro[1,6] and Sheerazed Boulkroun[1#]

[1] Université Paris Cité, Inserm, PARCC, F-75015 Paris France
[2] Electrophysiology Facility, University of Zurich, Zurich, Switzerland
[3] CIC1418 and DMU CARTE, Assistance Publique Hôpitaux de Paris (AP-HP), Hôpital Européen Georges Pompidou, F-75015 Paris, France
[4] Howard Hughes Medical Institute & Department of Neurosciences, University of California, San Diego, San Diego, United States.
[5] Service de Physiologie, Assistance Publique Hôpitaux de Paris (AP-HP), Hôpital Européen Georges Pompidou, F-75015 Paris, France
[6] Service de Génétique, Assistance Publique Hôpitaux de Paris (AP-HP), Hôpital Européen Georges Pompidou, F-75015 Paris, France

[#]Corresponding author

Address correspondence to:
Sheerazed Boulkroun, PhD
INSERM, U970
Paris Cardiovascular Research Center – PARCC
56, rue Leblanc,
75015 Paris – France
Tel : +33 (0)1 53 98 80 24
Fax : + 33 (0)1 53 98 79 52
e-mail : sheerazed.boulkroun@inserm.fr

**Running Title**: Molecular mechanisms responsible for PA

Word count – Abstract: 250
Word count – manuscript: 9880
Number of figures: 5
Number of tables: 1

**Abstract**

**Background**. Primary aldosteronism is the most common form of secondary hypertension. The most frequent genetic cause of aldosterone producing adenoma (APA) are somatic mutations in the potassium channel KCNJ5. They affect the ion selectivity of the channel, with sodium influx leading to cell membrane depolarization and activation of calcium signalling, the major trigger for aldosterone biosynthesis.

**Methods**. To investigate how *KCNJ5* mutations lead to the development of APA, we established an adrenocortical cell model in which sodium entry into the cells can be modulated "on demand" using chemogenetic tools (H295R-S2 α7-5HT3-R cells). We investigated their functional and molecular characteristics with regard to aldosterone biosynthesis and cell proliferation.

**Results.** A clonal cell line with stable expression of the chimeric α7-5HT3 receptor in H295R-S2 cells was obtained. Increased sodium entry through the α7-5HT3 receptor upon stimulation with uPSEM-817 led to cell membrane depolarization, opening of voltage-gated $Ca^{2+}$ channels and increased intracellular $Ca^{2+}$ concentrations, resulting in the stimulation of *CYP11B2* expression and increased aldosterone biosynthesis. Increased intracellular sodium influx did not increase proliferation, but rather induced apoptosis. RNA sequencing and steroidome analyses revealed unique profiles associated with $Na^+$ entry, with only partial overlap with angiotensin II or potassium induced changes.

**Conclusion.** H295R-S2 α7-5HT3-R cells are a new model reproducing the major features of cells harboring *KCNJ5* mutations. Increased expression of *CYP11B2* and stimulation of the mineralocorticoid biosynthesis pathway are associated with a decrease of cell proliferation and an increase of apoptosis, indicating that additional events may be required for the development of APA.

**Key words:** primary aldosteronism, adrenal cortex, aldosterone biosynthesis, calcium signaling, cell proliferation

**Nonstandard abbreviations**

ZG: Zona Glomerulosa; PSEM: Pharmacologically Selective Effector Molecules; PSAM: Pharmacologically Selective Actuator Modules; PA: Primary Aldosteronism; APA: Aldosterone Producing Adenoma

**Introduction**

The adrenal gland consists of two distinct regions: the outer adrenal cortex and the inner adrenal medulla. The adrenal cortex is further divided into three zones: the *zona glomerulosa* (ZG), *zona fasciculata*, and *zona reticularis*, each specialized in hormone production due to the expression of specific enzymes. Among these hormones, aldosterone, produced by the adrenal ZG, plays an important role in regulating salt and fluid balance, thereby controlling arterial blood pressure. As a key component of the renin-angiotensin-aldosterone system (RAAS), its production is mainly stimulated by angiotensin II (AngII), which increases in response to volume depletion, and elevated plasma potassium ($K^+$) levels[1,2]. AngII signals through its type 1 receptor (AT1R) and activates, via Gαq, an entire signaling cascades that leads to the release of $Ca^{2+}$ from the endoplasmic reticulum. The stimulation by $K^+$ and to a less extend by AngII results in ZG cell membrane depolarization and opening of voltage-gated calcium ($Ca^{2+}$) channels, leading to an increased intracellular $Ca^{2+}$ concentration. The activation of $Ca^{2+}$ signaling triggers a phosphorylation cascade that leads to increased transcription of *CYP11B2*, coding for aldosterone synthase, and aldosterone biosynthesis[1].

Deregulation of the mechanisms regulating adrenal aldosterone biosynthesis results in primary aldosteronism (PA). PA is the leading cause of secondary hypertension affecting approximately 5-10% of hypertensive patients and up to 20% of those with treatment resistant hypertension[3–6]. It is characterized by hypertension with elevated aldosterone levels, low plasma renin concentration, increased aldosterone to renin ratio and often associated with hypokalaemia. The main subtypes of PA are bilateral adrenal hyperplasia and aldosterone-producing adenomas (APA), accounting together for 95% of cases. PA is associated with an increased risk of cardiometabolic and renal complications[7] due to the major adverse effects of aldosterone excess; therefore, early diagnosis and appropriate treatment of PA are essential.

In the past decade, research has uncovered mutations in ion channels (*KCNJ5*[8], *CACNA1D*[9,10], *CACNA1H*[11,12], *CLCN2*[13,14], *SLC30A1*[15]) and pumps (*ATP1A1*[9,16], *ATP2B3*[16]), as principal causes of APA and familial forms of PA. These mutations enhance either directly or indirectly intracellular $Ca^{2+}$ concentrations, the main trigger for aldosterone biosynthesis. Notably, in the case of *KCNJ5*, which encodes the G protein-coupled inwardly rectifying $K^+$ channel (GIRK4), the majority of mutations cluster near the channel's ion-selective filter, leading to a loss of selectivity for $K^+$ ions in favour of an intracellular sodium ($Na^+$) influx[8]. These mutations lead to cell membrane depolarization, the opening of voltage-gated $Ca^{2+}$ channels, an increase of intracellular $Ca^{2+}$ concentration, activation of $Ca^{2+}$ signalling pathways, and ultimately an increase of *CYP11B2* expression and aldosterone biosynthesis[8].

While the role of *KCNJ5* mutations in autonomous aldosterone production has been well established, their role in modulating cell proliferation is still under debate[17–20]. The first objective of our study was therefore to investigate the role of *KCNJ5* mutations in regulating proliferative processes that could lead to adenoma development. The second objective was to determine whether changes in intracellular sodium balance could induce specific intracellular transcriptional profiles that could explain the peculiar pathological and biological features of adenoma carrying *KCNJ5* mutations. For this purpose, we used chemogenetic tools which allow to manipulate specific ion fluxes via modified ion channels with pharmacologically selective properties, known as Pharmacologically Selective Actuator Modules (PSAM). PSAM, comprising mutated ligand-binding domains and selective ionic pore domains respond to Pharmacologically Selective Effector Molecules (PSEM), inducing channel opening and allowing specific ions to flow through the activated channel[21,22]. Here we employed second-generation chemogenetic tools to established a human adrenal cell model in which we could modulate sodium entry on demand, by introducing the α7-5HT3 chimeric receptor. This chimeric receptor consists of the ligand binding domain of the α7 nicotinic acetylcholine receptor and the ionic pore domain of the serotonin receptor type 3. Importantly, the ligand binding domain has been mutated to bind only the PSEM and not the endogenous ligand[21]. Activation of this α7-5HT3 receptor (α7-5HT3-R) by uPSEM-817 was used to study the effect of $Na^+$ entry into the cells, mimicking molecular abnormalities observed in the presence of *KCNJ5* mutations. To determine the functional and molecular characteristics of this new cell model, we assessed changes in membrane potential, intracellular $Ca^{2+}$ concentration, and impact on cell proliferation. Additionally, we conducted RNA sequencing and steroidomic analyses following treatment with uPSEM-817, AngII, and $K^+$.

**Material and Methods**

**Data Availability**

An expended Methods section is available in the online-only Data Supplement.

The data, analytic methods and study materials that support the findings of this study are available from the corresponding author upon reasonable request.

*<u>Cell culture and electroporation</u>*

The H295R-S2 cell line, a subclone of the H295R human adrenocortical carcinoma cell, was kindly provided by Dr. William E. Rainey[23] and cultured in complete medium containing DMEM/Eagle's F12 medium (1:1) (GIBCO, Life Technologies) supplemented with 2% Ultroser G (Sartorius, France), 1% insulin/transferrin/selenium premix (BD Bioscineces), 7.5 mM HEPES (GIBCO), 1% penicillin and streptomycin (GIBCO, Life Technologies) and 20 mg/mL G418 (Thermo Fisher Scientific). Cells were maintained at 37°C under a humid atmosphere of 95% air and 5% $CO_2$.

The α7-5HT3 chimeric receptor consists of the ligand binding domain of the nicotinic receptor fused with the ionic pore domain of the serotonin receptor type 3. The sequence has been inserted into pcDNA3.1 vector[21].

5.000.000 of H295R-S2 cells were seeded into a 100mm tissue culture dish. After 24h cells were tripsinized, counted and $3.10^6$ cells resuspended in 100μl Nucleofactor R solution and electroporated with 3μg of plasmid (pcDNA3 containing or not α7-5HT3 cDNA) using the Amaxa nucleofactor kit R (Lonza) according to the manufacturer instructions. After electroporation, the mixed populations were amplified in under selection pressure with G418. Pure clones were isolated by picking clones after limited dilution of the cells in order to isolate one cell at an optimum distance from the other in a 200mm diameter plate. Colonies were, afterwards, isolated in wells and amplified for further characterization.

*<u>Statistics</u>*

The number of independent experiments (n) refer to the number of cells or dishes studied to calculate mean±SEM or medians. The measurements were carried out at different days and from different cell preparations using different cell passages to ensure the reproducibility of the experiments. For patch-clamp experiments, each cell was analyzed separately.

Data were analysed in Prism10 software (GraphPad, San Diego, CA) using the appropriate statistical tests as indicated in the text. The normality of the data distribution was checked using the ShapiroWilk test. Quantitative variables were reported as means±s.e.m. when a Gaussian distribution was present or as medians and interquartile ranges when no Gaussian distribution was present. Pairwise comparisons were conducted using unpaired t tests for normally distributed data and Mann–Whitney tests for non-normally distributed data. For multiple comparisons, ANOVA followed by Bonferroni was applied when data presented a Gaussian distribution while Kruskal–Wallis followed by Dunn's test was used for non-Gaussian distribution was present. A *p* value < 0.05 was considered significant.

**Results**

*Characterization of a cell model expressing the α7-5HT3 receptor*

To assess the impact of modulating intracellular Na$^+$ concentration on adrenal cell function, we established stable expression of the chimeric α7-5HT3-R in H295R-S2 cells. This receptor was formed by combining the extracellular ligand-binding domain of the α7 nicotinic acetylcholine receptor and the ion pore domain of the serotonin receptor 5HT3[21]. Expression of the α7-5HT3-R was investigated by RT-qPCR[24] and found exclusively in cells transfected with the vector containing its genetic sequence (Figure S1A), but not in control cells transfected with an empty vector. Upon treatment with 12mM of K$^+$ both cell lines exhibited a rapid increase in intracellular Ca$^{2+}$ concentration (Figure S1B). However, treatment with different concentrations of varenicline, a nicotinic receptor partial agonist and cholinergic agonist, resulted in a dose-dependent increase in intracellular Ca$^{2+}$ concentration in cells expressing the α7-5HT3-R only (Figure S1B), indicating a specific effect on the chimeric receptor. This was associated with an increase in *CYP11B2* mRNA expression for varenicline concentrations ranging from $10^{-8}$ to $10^{-5}$M (Figure S1C).

To further characterize the cells expressing the α7-5HT3-R, monoclonal cell populations were obtained via limiting dilution under antibiotic selection pressure. Among the 19 monoclonal cell populations selected, six showed α7-5HT3-R mRNA expression (Figure S1D), and in-depth characterization was carried out on three of them, designated as clones 17, 24, and 42 (Figure 1). The expression of α7-5HT3-R mRNA was not statistically different between clones 17, 24, and 42 (Figure 1A).

Brightfield images revealed no discernible variations in cell morphology between control cells and the three selected clones (Figure 1B). Examination of the cellular localization of the α7-5HT3-R using α-bungarotoxin, a peptide binding to the α subunit of the nicotinic acetylcholine receptor, showed its presence on the cell surface of clones 17, 24, and 42 but not in control cells (Figure 1B). To ensure that the expression of the α7-5HT3-R did not induce structural changes, we studied the expression of markers of cytoskeletal organisation and of ZG cells. Phalloidin staining revealed similar actin organization in cells expressing an empty vector (control cells) or expressing the α7-5HT3-R (Figure 1B). A similar staining was observed for Disabled-2 (Dab2) (Figure 1B), a protein marker of ZG cells expressed at the cell membrane.

Varenicline is a nicotinic receptor partial agonist and a cholinergic agonist, which may have non-specific effects in our cell model. To avoid this, we compared the effect of varenicline to those of a drug, uPSEM-817, specifically designed to bind to the α7-5HT3-R. Similar responses were obtained in response to uPSEM-817 and varenicline treatment in intracellular $Ca^{2+}$ responses, both as traces (Figure S2A, S2C) and maximum of activation (Figure S2B, S2D).

### *α7-5HT3 receptor activation by PSEM-817 induces cell membrane depolarization and increases intracellular $Ca^{2+}$ content via $Na^+$ influx*

Perforated patch-clamp recordings were conducted to assess the electrophysiological properties of two of the three clones expressing the α7-5HT3-R and selected for further investigations (clones 17 and 42) in comparison to control cells. Control cells expressing an empty vector, as well as clones 17 and 42, displayed similar membrane hyperpolarization with membrane potentials of -60.83±1.2, -61.73±1.16, and -63.27±1.26, respectively (Figure 2A), indicating that the α7-5HT3-R was not leaking ions.

All the clones responded to induction by AngII, exhibiting membrane depolarization upon the application of $10^{-8}$M AngII (ΔEm=48.26±3.52, 51.81±3.73, 46.00±3,51 for control cells, clones 17 and 42 respectively) (Figure 2B). However, only clones 17 and 42 were responsive to $10^{-7}$M ($\Delta$Em=41.58±4.12 and 41.49±2.13 for clones 17 and 42 respectively) and $10^{-5}$M of uPSEM-817 ($\Delta$Em=49.03±5.51 and 47.19±2.24 for clones 17 and 42 respectively). The lower concentration of $10^{-9}$M uPSEM-817 did not consistently induce membrane depolarization (Figure 2B).

Elecrophysiological measurements were performed using High Throughput Automated Patch-Clamp to determine whether activation of the $\alpha$7-5HT3-R by uPSEM-817 increased Na$^+$ current. Repetitive stimulations using a voltage ramp from -100mV to +20mV that was used to calculate the reversal potential (ERev) (Figure S3A), a surrogate of the resting membrane voltage, confirmed that uPSEM-817 induced depolarization of clones 24 and 42 compared to vehicle treated cells (Figure S3B). Inward currents were measured at -100mV as they correspond to positive charges (e.g. Na$^+$ ions) entering the cells. Treatment with $10^{-6}$M of uPSEM-817 under extracellular Na$^+$ concentrations of 80mM or 132.5mM activated a statistically significant inward current in clone 42 and a similar trend in clone 24 (Figure S3C). These results are consistent with an increase of Na$^+$ permeability when cells were treated with uPSEM-817.

To determine whether the membrane depolarization triggered by uPSEM-817 correlated with an increase in intracellular Ca$^{2+}$ concentration, $\alpha$7-5HT3-R expressing cells were exposed to concentrations of uPSEM-817 ranging from $10^{-9}$ to $10^{-5}$M, as well as 12mM K$^+$ and $10^{-8}$M AngII as positive controls (Figure 2C-E). The results are presented for clone 17 (Figure 2C-E) and similar results were obtained for clone 42 (Figure S4). Application of AngII or K$^+$ led to a rapid increase in intracellular Ca$^{2+}$ levels followed by a rapid decline without returning to baseline levels (Figure 2C, Figure S4A), occurring after similar response latency (data not shown). The maximum of activation (Figure 2D, Figure S4B) was significantly increased when cells were treated with AngII or $K^+$. The area under the curve (AUC) at the peak (between 0 and 360s) (Figure 2E, left panel, Figure S4C) and at the steady state response (between 0 and 700s) (Figure 2E, right panel, Figure S4D) were also significantly increased after $K^+$ or AngII treatment, these increases being more important when cells were treated with $K^+$. Treatment with uPSEM-817 ranging from $10^{-9}$ to $10^{-5}$M also resulted in a similar rapid increase in intracellular $Ca^{2+}$ content; this increase was more modest at $10^{-9}$M uPSEM-817 compared to higher concentrations of the compound. Interestingly, the decline following the peak was slower with uPSEM-817 compared to AngII or $K^+$ (Figure 2C, Figure S4A). Maximum activation (Figure 2D, Figure S4B) and AUC at the peak (Figure 2E, left panel, Figure S4C) and at steady state (Figure 2E, right panel, Figure S4D) were lower in response to $10^{-9}$M of uPSEM-817 than with $10^{-7}$ and $10^{-5}$M uPSEM-817. Despite variations in $Ca^{2+}$ content, while the maximum activations were comparable when cells were treated with AngII, $K^+$, $10^{-7}$ and $10^{-5}$M uPSEM-817, the AUC at the steady state were significantly higher when cells were treated with $10^{-7}$ and $10^{-5}$M uPSEM-817 compared to AngII and $K^+$. The sustained higher $Ca^{2+}$ content when cells were treated with PSEM suggests the activation of different mechanisms compared to AngII and $K^+$.

*α7-5HT3 receptor activation by uPSEM-817 activates steroidogenesis*

The steroidome of clones 17, 24 and 42 was determined using the cell supernatant after 8h (Figure S5A-B) or 24h (Figure S5C-D) of treatment with $10^{-8}$M AngII or 12mM $K^+$ as positive controls (Figure S5A, S5C), and uPSEM-817 ranging from $10^{-9}$ to $10^{-5}$M (Figure S5B, S5D), and the results presented separately for each clone (Figure S5). Interestingly, despite similar modification in intracellular $Ca^{2+}$ concentration, different profiles were obtained for the three different clones in response to AngII, $K^+$ and uPSEM-817. However, they converge all towards a stimulation of steroid biosynthesis in response to uPSEM-817. Among the 19 steroids measured, only 14 were detectable after 8h and 24h of treatment. After 8h of treatment, an overall activation of steroid biosynthesis was observed in response to AngII and $K^+$ (Figure S5A, Table 1). Similarly, uPSEM-817 at concentrations from $10^{-7}$ to $10^{-5}$M led to an increase in steroid biosynthesis, with no significant changes observed at $10^{-9}$ and $10^{-8}$M of uPSEM-817 (Figure S5B, Table 1). Interestingly, the increases in certain steroids due to uPSEM-817 were less pronounced compared to those induced by AngII and $K^+$. After 24h of treatment, a significant decrease in the concentration of two steroid precursors, pregnenolone and progesterone, was observed in response to AngII and $K^+$, while DOC, 17-hydroxyprogesterone, 17-hydroxypregnenolone, pregnenolone and progesterone concentrations remained elevated in response to uPSEM-817 (Table 1). Aldosterone biosynthesis was highly stimulated by AngII and $K^+$ and to a lesser extent by uPSEM-817. While no effect was observed after 8h of treatment, after 24h the concentration of 18-oxocortisol was significantly increased in response to $K^+$, and a trend was also observed in response to AngII and for all the concentrations of uPSEM-817. The concentrations of 18-hydroxycortisol were undetectable in all tested conditions.

### *T-type and L-type channels are both involved in $Ca^{2+}$ entry into the cells in response to uPSEM-817*

Both T-type and L-type voltage-gated $Ca^{2+}$ channels have been shown to be involved in aldosterone biosynthesis regulation by modulating intracellular $Ca^{2+}$ concentrations[25,26]. We evaluated the effect of a 2h pre-treatment with $10^{-6}$M of the L-type $Ca^{2+}$ channel blocker nifedipine, or $10^{-6}$M of the T-type $Ca^{2+}$ channel blocker mibefradil on the modulation of intracellular $Ca^{2+}$ content in response to $10^{-8}$M AngII, 12mM $K^+$ and increasing concentrations of uPSEM-817 (Figure 3A-C). Interestingly, pre-treatment of the cells with nifedipine or mibefradil partially abolished $Ca^{2+}$ entry into the cells after treatment with $K^+$ and uPSEM-817, as revealed by lower peak (Figure 3A-B) and decreased AUC (representing the modifications of intracellular $Ca^{2+}$ concentrations, illustrating activation of $Ca^{2+}$ signaling) (Figure 3C). In contrast, nifedipine and mibefradil had no effect on the initial entry of $Ca^{2+}$ into the cell in response to AngII (Figure 3B), and only nifedipine pre-treatment led to a significant decrease in the AUC (Figure 3A-C). These results indicate that both T- and L-type $Ca^{2+}$ channels are mobilized to allow $Ca^{2+}$ entry into the cells in response to $K^+$ and uPSEM-817, a mobilization that appears to be greater than in response to AngII.

To determine if the differences observed in the modulation of intracellular $Ca^{2+}$ concentration in response to $Ca^{2+}$ channel blockers and uPSEM-817 cotreatment were due to differences in cell membrane depolarization, we performed patch-clamp analyses (Figure 3D). Prior to the administration of $10^{-7}$M uPSEM-817, we treated each patched cell with $10^{-6}$M mibefradil, or with $10^{-6}$M Nifedipine. These blockers were applied locally using an ejector solution. The results displayed a range of responses among cells: in certain instances, the blockers effectively suppressed depolarization, returning the membrane potential to a state near its baseline ($\Delta EM \approx 0$); while in other cases, the membrane remained depolarized, albeit to a lesser degree than when exposed exclusively to $10^{-7}$M uPSEM-817 (Figure 3D).

### *α7-5HT3 receptor activation by uPSEM-817 led to a decrease in cell proliferation and an increase of apoptosis*

To evaluate the impact of elevated intracellular $Ca^{2+}$ concentration on cell proliferation, cells were treated with 12mM $K^+$ or $10^{-9}$ to $10^{-5}$M uPSEM-817 for 24h and 72h. The number of viable proliferating cells was determined using a colorimetric method. Whereas following 24h of treatment, cell proliferation remained unaffected by $K^+$ or uPSEM-817 treatment, a decrease in cell proliferation was observed after 72h of treatment with the higher concentration of uPSEM-817 (Figure 4A).

To explore the impact on cell cycle progression of AngII, $K^+$ and uPSEM-817 treatment, we assessed the cell cycle phase distribution using propidium iodide staining and cytometric analysis at 8h, 24h, 48h, and 72h. Treatment with AngII for 8h resulted in a reduction in the percentage of cells in the G1 phase and an increase in the percentage of cells in G2 compared to untreated cells suggesting an increase in cell proliferation (Figure 4B). After 24h of AngII treatment, an increase in the percentage of cells in the G1 phase and a decrease in the S phase was also observed. Longer exposure to AngII had no effect on the distribution of cells across different phases of the cell cycle. In response to $K^+$ treatment, a similar pattern was observed with a decrease in the percentage of cells in G1, and an increase in G2 after 8h. While no alterations in cell distribution were noted after 24h of K+ treatment, an increase in the percentage of cells in G1 and a decrease in G2 was observed (Figure 4B). Conversely, treatment with uPSEM-817 did not affect cell distribution among the different phases compared to untreated cells (Figure 4C). We assessed the effect of uPSEM-817 on apoptosis by determining the proportion of cells in the sub-G1 phase. The sub-G1 phase reflects DNA fragmentation which occurs in the late stage of apoptosis. Similar to what was observed for cell proliferation results, no effect was observed when cells were treated with uPSEM-817, AngII and $K^+$ for 24h (Figure 4D). At 72h, a significant increase in the percentage of cells in sub-G1 phase was observed in response to $K^+$, AngII and uPSEM-817 ($10^{-7}$M) (Figure 4E), indicating an increase in cell apoptosis.

*α7-5HT3 receptor activation by uPSEM-817 results in the activation of specific pathways*

To gain insight into the impact of modulating intracellular $Na^+$ concentration in cells expressing the α7-5HT3-R, we conducted RNA sequencing analysis on these cells (clones 17, 24, and 42) following treatment with $10^{-7}$M uPSEM-817 for 8h (Figure S6 and S7) and 24h (Figure 5, S8). Hierarchical clustering effectively segregated cells treated for 8h and 24h from untreated cells (Figure S6A, S7A, S8A, 5A). After 8h of treatment, we identified 28 differentially expressed genes with a fold change of at least 0.5; 18 upregulated genes (64.29%) and 10 downregulated genes (35.71%) (Table S2). Gene ontology analyses unveiled 18 specific enriched biological processes in uPSEM-817 treated cells (Figure S6B, Table S3), primarily related to $Ca^{2+}$ ion transport and signalling pathways, cell adhesion, and aldosterone synthesis and secretion. Following 24h of treatment, 30 genes were differentially expressed with a fold change of at least 0.5, with 22 genes (73.33%) upregulated and 8 (26.67%) downregulated (Figure 5B, Table S4)[27,28]. Enriched pathways included protein regulation of GTPase activity, Ras protein signal transduction, and aldosterone synthesis and regulation (Figure 5B, Table 3). These analyses were completed by using Gene Set Enrichment Analysis (GSEA) and Hallmark database that provides information on most universal cellular mechanisms (Figures S7, S8). Using an FDR <25% we identified 6 differentially regulated pathways after 8h of treatment with PSEM and 7 after 24h. After 8h of treatment, TNFα signaling via NFκB and coagulation pathway were found enriched in basal condition (Figure S7B) whereas Kras signaling, heme metabolism, UV response and Interferon alpha response were found to be enriched in response to uPSEM-817 (Figure S7C). After 24h of treatment (Figure S8), only TNFα signaling via NFκB pathway (Figure S8B) was enriched in basal condition and Myc targets, oxidative phosphorylation, DNA repair Apical surface, unfolded protein response and G2M checkpoint pathways were enriched after uPSEM-817 treatment (Figure S8C).

To identify genes and pathways specifically regulated by modulation of intracellular $Na^+$ concentration, we also performed RNA sequencing on α7-5HT3-R expressing cells treated with 12mM $K^+$ and $10^{-8}$M AngII for 8h and 24h (Figure S6, S9, Figure 5). After 8h of treatment, 4932 genes were differentially expressed in response to AngII and 728 in response to $K^+$ (Figure S6C, S6G, Table S6, Table S7). Gene ontology analyses[27,28] for 8h AngII treatment revealed enrichment in 390 different pathways (Table S8), with significant involvement in the regulation of reticulum activity, negative regulation of cell growth, and aldosterone, cortisol and parathyroid hormone synthesis and secretion (Figure S6D). Eight hours of treatment with $K^+$ resulted in enrichment of 211 pathways (Table S9), including membrane depolarization during action potential, $Ca^{2+}$ and $Na^+$ ion transport, and circadian rhythm (Figure S6H). Interestingly, both AngII and $K^+$ treatment led to enrichment in positive regulation of transcription and MAPK signaling pathways. Among the 28 differentially expressed genes in response to uPSEM-817, 17 were common to AngII (*OLFML2B*, *RGPD6*, *LMOD1*, *MAT1A*, *MC2R*, *CA2*, *CACNA1C*, *EGFR*, *ETV5*, *TNXB*, *PFKP*, antisense to *WASF3*, antisense to *ERVW-1* and *PEX1*, *PPP2R1B*, *LYPLAL1-AS1*, *LONRF2* and *ST6GALNAC6*) (Figure S6E) and 12 to $K^+$ (*OLFML2B*, *LMOD1*, *MAT1A*, *MC2R*, *CA2*, *CACNA1C*, *ATP8B4*, *ETV5 PFKP*, *PPP2R1B*, *LYPLAL1-AS1* and *LONRF2* (Figure S6I). Among the 18 enriched pathways in response to uPSEM-817, 9 were commonly enriched in response to AngII (Figure S6F) and 5 in response to $K^+$ (Figure S6J). Most importantly, 11 genes were specifically regulated by uPSEM-817 when compared with AngII (*HPX*, *EPHA10*, *ATP8B4*, *HHIP*, *MICAL1*, *RBMS2*, *E9PAM4*, *H7C2Y5*, *TP53TG3B*, *DUXAP8* and *RNVU1-7*) and 16 when compared to $K^+$ (*HPX*, antisense to *WASF3*, *EPHA10*, *HHIP*, *ST6GALNAC6*, antisense to *ERVW-1* and *PEX1*, *EGFR*, *MICAL1*, *TNXB*, *RGPD6*, *RBMS2*, *E9PAM4*, *H7C2Y5*, *TP53TG3B*, *DUXAP8* and *RNVU1-7*). This allowed us to define a specific $Na^+$-induced specific signature composed of 10 genes (*HPX*, *EPHA10*, *HHIP*, *MICAL1*, *RBMS2*, *E9PAM4*, *H7C2Y5*, *TP53TG3B*, *DUXAP8* and *RNVU1-7*) (Figure S10).

After 24h of treatment, 589 genes were statistically differentially expressed in response to AngII (Figure 5C, 5D, Table S10). Among them, 12 were common to uPSEM-817 (*SOBP*, *ADRA2A*, *ETV5*, *antisense to DLGAP1*, *VSIR*, *GCNA*, *FCHSD1*, *RAPGEFL1*, *PPP2R1B*, *CYP11B2*, *LYPLAL1-AS1* and *PRSS53*); 146 pathways were specifically enriched in response to AngII and 2 (Cushing syndrome and Aldosterone synthesis and secretion) were commonly enriched in response to AngII and uPSEM-817 (Figure S9B, Table S11). Interestingly, enriched pathways included those associated with cell adhesion and extracellular matrix organization (Figure S9A). $K^+$ treatment for 24h led to the regulation of 798 genes (Figure S7A, 5F, Table S12), among them 12 were common to uPSEM-817 (*KLF10*, *ADRA2A*, *C9ORF72*, *MC2R*, *ETV5*, *RAPGEF4*, *VSIR*, *GCNA*, *FCHSD1*, *PPP2R1B*, *CYP11B2* and *LYPLAL1-AS1*). Similarly, enriched pathways included those associated with cell adhesion and extracellular matrix organization (Figure S9C, Table S13). Five out of the eight enriched pathways identified in response to uPSEM-817 treatment were also commonly enriched in response to $K^+$ (Figure S9D). These common pathways involved positive regulation of GTPase activity, Cushing Syndrome, Negative regulation of cell migration, Ras protein signal transduction, and Aldosterone synthesis and secretion. Notably, Cushing syndrome and aldosterone biosynthesis and secretion pathways were also commonly enriched between uPSEM-817 and AngII. We were able to define a 24h sodium-specific signature composed by 14 genes (*FMNL1-AS1*, *IDI2-AS1*, *WHAMMP1*, *MUSTN1-ITIH4 Readthrough*, *NFKBIZ*, *KCNQ1OT1*, *SH3BP5-AS1*, CDK10, *PTPRU*, *IL17RD*, *TMEM67*, *MAZ-AS*, *DOP1B* and one unknown sequence) (Figure S11).

The list of differentially expressed genes was compared to their expression in 11 control adrenal and 123 APA retrieved from transcriptomic data[29] (Table S14, S15). Of these 123 APA, 50 carried a mutation in the *KCNJ5* gene and 73 in another APA driver gene; the comparison was therefore made between the 11 control adrenals, the 50 APA with a mutation in the *KCNJ5* gene and the 73 APA without a mutation in the *KCNJ5* gene. After 8h of treatment with $10^{-7}$M of uPSEM-817, among the 55 identified genes that were significatively differentially expressed without consideration of the fold change, the expression of 49 were found in our transcriptomic data from control adrenals and APA and could be retrieved. In these data, no expression was detected for five genes, and out of the remaining 44 genes, 13 were also found to be differentially expressed in APA compared to control adrenal (Table S12), two only in APA with a KCNJ5 mutation (*HPX* and *WHRN*) and three in APA without KCNJ5 mutation (ETV5, *GPAM* and *OLFML2B*). Interestingly, among the 13 genes differentially expressed in APA independently of the mutational status, 4 were found to be up-regulated (*CACNA1C*, *ATP2B2*, *VDR* and *EPHA10*) and 1 downregulated (*RGPD6)* in both APA and α7-5HT3-R cells, whereas 6 genes were found to be upregulated in α7-5HT3-R cells but downregulated in APA (*PPP2R1B*, *LONRF2*, *ATP8B4*, *TNXB*, *ITGA7* and *HSPA12A*), independently of the mutational status and 2 down-regulated in response to uPSEM-817 but up-regulated in all APA (*TMEM200A* and *ELL2*). Similarly, after 24h of treatment, among the 73 genes identified genes that were significantly differentially expressed regardless of the fold change, the expression of 64 genes could be retrieved from our transcriptomic data. No expression was found for 10 genes, and among the remaining 54 genes, 15 were also found to be differentially expressed in APA compared to control adrenal (Table S13). Among these 15 genes, three were found to be commonly upregulated in both α7-5HT3-R cells and APA (*FCHSD1*, *CACNA1C* and *TTN1*) and one downregulated (*KLF10*) whereas two were found to upregulated in α7-5HT3-R cells and downregulated in APA, independently of the mutational status (*PPP2R1B* and *NEAT1*) and four downregulated in α7-5HT3-R cells and upregulated in all APA (*STMN3*, *FIBCD1*, *FSCN1* and *CCDC71L*). Interestingly, three genes were found to be significantly regulated only in APA without KCNJ5 mutations, two downregulated in α7-5HT3-R cells but upregulated in APA (*ETV5* and *CEP170*) and one downregulated in both (*NR2F2*); two genes were found to be significantly regulated only in APA with KCNJ5 mutations, one downregulated in α7-5HT3-R cells but upregulated in APA (*NFKBI2*) and one downregulated in both (*ABHD2*). Interestingly, none of the genes commonly regulated in α7-5HT3-R cells and APA belongs to the Na$^+$ signature defined after 8h and 24h of treatment suggesting that they are probably involved in the early phase of APA development.

**Discussion**

Somatic *KCNJ5* mutations are the most frequent genetic abnormalities found in APA with a prevalence between 43% and 75% of cases. APA with *KCNJ5* mutations are more frequent in women and at younger age[29,30] and are characterized by hybrid steroid production and larger adenoma size[31]. Expression of mutated *KCNJ5* in adrenocortical cells leads to increased aldosterone production without increasing cell proliferation[17,19], raising the question as to whether those mutations are sufficient to lead to both increased aldosterone production and adenoma formation. Here we describe the development of an adrenocortical cell model that recapitulates the main features of *KCNJ5* mutations, by modulating "on demand" sodium entry into the cells using chemogenetic tools. Increased sodium entry through the chimeric α7-5HT3-R upon stimulation with uPSEM-817 led to cell membrane depolarization, opening of voltage-gated $Ca^{2+}$ channels and increased intracellular $Ca^{2+}$ concentrations, resulting in the stimulation of *CYP11B2* expression and increased aldosterone biosynthesis.

The steroid response of H295R-S2 α7-5HT3-R cells to AngII is similar to that described in H295R cells[32], with an early and transitory increase of pregnenolone, progesterone and 11-deoxycorticosterone and a late and sustained production of aldosterone and corticosterone[32]. In contrast, sodium entry into the cells induced by uPSEM-817 led to a specific steroid profile only partially overlapping with that induced by AngII and $K^+$. In particular, we observed a prolonged induction of early steroid precursors and DOC, suggesting an action on both early and late steps of aldosterone biosynthesis, as well as a delayed and smaller induction of aldosterone biosynthesis; there was no induction of glucocorticoid biosynthesis in H295R-S2 α7-5HT3-R cells. An increase of 18-oxocortisol and 18-hydroxycortisol was also observed in response to uPSEM-817 in two of the three clones, suggesting production of hybrid steroids in response to increased intracellular $Na^+$ concentration. This is consistent with results reported for the overexpression of KCNJ5 harboring the p.Tyr158Ala mutation in HAC cells resulting in a significant increase of both 18-hydroxycortisol and 18-oxocortisol[18,33]. However, despite similar depolarization of the cells in response to uPSEM-817 or AngII, our data suggest a delayed activation of mineralocorticoid biosynthesis in response to the modulation of intracellular $Na^+$ concentration compared to AngII and $K^+$ that could be due to the activation of specific signaling pathways. This hypothesis is supported by our finding of a specific gene expression signature associated with $Na^+$ influx into the cells.

Interestingly, gene expression analysis revealed expression of *KCNJ5* in H295R-S2 α7-5HT3-R cells (Figure S12A), which remains unchanged after uPSEM-817 treatment. In this context, $Na^+$ may act as an activator for the KCNJ5 channel[34] by binding to the C-terminal part of the channel, near a region also sensitive to the $PtdIns(4,5)P_2$, enhancing the sensitivity of the channels to $PtdIns(4,5)P_2$[35]. The interaction of $PtdIns(4,5)P_2$ with the C-terminal part of GIRK4 has been shown to regulate its opening by stabilizing the structure of the pore[35]. The resulting $K^+$ extrusion may attenuate the activation of $Ca^{2+}$ signaling and explain the delayed increased in aldosterone biosynthesis observed in response to uPSEM817.

If the role of *KCNJ5* mutations in inducing aldosterone biosynthesis has been clearly established, their role in promoting abnormal cell proliferation is still a matter of debate. In our model, we demonstrated that $Na^+$ entry into the cells lead to decreased cell proliferation and increased apoptosis. This is in accordance with previous studies showing that expression of *KCNJ5* mutants in adrenocortical cells resulted in decreased cell proliferation associated in some cases with increased apoptosis[18,20]. *Choi et al* postulated that the activation of $Ca^{2+}$ signaling induced by *KCNJ5* mutations is responsible for both[8] increased aldosterone biosynthesis and cell proliferation, and indeed some studies also reported an association between *KCNJ5* mutations and larger adenoma size[29,30] and a positive correlation between expression of Ki67, a marker of cell proliferation, and the diameter of APA harboring *KCNJ5* mutations[20]. Specific factors overexpressed in APA harboring *KCNJ5* mutations[36,37] have been suggested to possibly counteract the pro-apoptotic effect of these mutations[20] suggesting, the existence of compensatory mechanisms maintaining cell proliferation over long term. In particular, expression of Teratocarcinoma-Derived Growth Factor-1 (*TDGF-1*) and Vinsinin-like 1 (*VSNL1*), two genes with anti-apoptotic properties, were found to be upregulated in *KCNJ5* mutated APA[38,39]. However, in our cell model the expression of *VSNL1* was not modified (Figure S12B) and *TDGF1* was not expressed at all. The increased expression of these genes may be a later event secondary to the development of an APA to compensate for $Na^+$-induced cell death. Moreover, signaling pathway analyses revealed downregulation of genes involved in the G2/M checkpoint pathway and in the progression through the cell division cycle, probably contributing to the decrease of cell proliferation observed in our model. Alternatively, our data suggest that other mechanisms may be responsible for abnormal cell proliferation, leading to the development of APA. These include the presence of two hits, one responsible for cell proliferation and the other for autonomous aldosterone production. Recently, we and others identified risk alleles associated to PA in genome wide association studies. Within the identified genetic loci, some genes appear to modulate adrenal cortex homeostasis and may affect cell proliferation, eventually generating a propitious environment leading to the occurrence of somatic mutations[40,41]. The two models are not mutually exclusive and may be linked together by mechanisms that remain to be identified. Finally, in FH-III it cannot be ruled out that $Na^+$ entry induced by *KCNJ5* mutations may have an impact on cell proliferation during adrenal development, which could explain the bilateral hyperplasia observed in severe cases[8,19].

AngII and potassium both stimulate aldosterone biosynthesis by activating $Ca^{2+}$ signaling, but through different mechanisms, resulting in only a partial overlap in the genes they induce. Similarly, uPSEM-817, by increasing intracellular $Na^+$ levels, induces membrane depolarization, which like potassium, leading to opening of voltage-gated $Ca^{2+}$ channels and increase of intracellular $Ca^{2+}$ concentration. However, uPSEM-817 also has unique effects, which account for the partial overlap in the differentially expressed genes induced by these three compounds. Gene expression profiles allowed us to identify a sodium-induced gene signature composed of 10 genes after 8h of treatment with uPSEM-817 (Figure S9) and 14 genes after 24h of treatment (Figure S10). Interestingly, among these lists of genes, some are involved in cell cycle regulation, proliferation and apoptosis. Hence, the expression of *RBMS2* and *PTPRU*, two genes involved in the control of cell cycle progression[42] and cell proliferation[43,44] respectively, was decreased after 8h or 24h while the expression of *CDK10*, a Cdc2-related kinase involved in the regulation of the G2/M phase of the cell cycle, was increased. However, the role of CDK10 in regulating cell proliferation is not clear, as some studies suggest a role of CDK10 in cell proliferation activation[45–47], while others report a tumor suppressor role for CDK10 through inhibition of cell proliferation[48,49]. The identification of these genes in the $Na^+$-signature supports our results showing an inhibitory effect of uPSEM-817 on cell proliferation. Among the genes belonging to the $Na^+$ signature after 8h of treatment with uPSEM-817, *HPX* expression was found to be upregulated. *HPX* encodes hemopexin, a protein with high binding affinity for heme. Interestingly, CYP11B2 and other cytochromes P450 enzymes use heme as cofactor required for their activity. The availability of heme has been shown to affect aldosterone and corticosterone biosynthesis in rats[50]. Our transcriptomic data revealed a significant increase of *HPX* expression in APA harboring a *KCNJ5* mutation[29] and a recent study reported also increased expression of *HPX* associated with CpG hypomethylation[51] in APA compared to adjacent adrenal gland suggesting a role of *HPX* in APA development potentially through the regulation of aldosterone biosynthesis. Similarly, the expression of *MC2R*, the melanocortin 2 receptor, was found to be increased in APA associated with CpG hypomethylation[51]. Gene expression profiling of H295R-S2 α7-5HT3-R cells treated with either AngII, K⁺ or uPSEM-817 revealed also a significant increase of *MC2R* expression; and a previous study reported higher expression of *MC2R* in APA compared to control adrenals[52]. Finally, a recent work of our laboratory shows that expression of MC2R in APA was more frequently found in regions expressing CYP11B2[53], suggesting a potential role in regulating aldosterone biosynthesis. Moreover, the existence of a continuum of PA and dysregulated aldosterone production, prominently influenced by ACTH, has recently been described[54].

In conclusion, we demonstrate that H295R-S2 α7-5HT3-R cells are a new cell model in which intracellular $Ca^{2+}$ concentration can be modulated on "demand", reproducing the major features of cells harbouring *KCNJ5* mutations. The stimulation of Na⁺ entry into the cells leads to cell membrane depolarization, $Ca^{2+}$ entry into the cells, activation of $Ca^{2+}$ signaling, increased expression of *CYP11B2* and stimulation of the mineralocorticoid biosynthesis pathway. This is associated with a decrease of cell proliferation and an increase of apoptosis, indicating that additional events may be required for the development of APA. RNA sequencing revealed that Na⁺ entry into the cells is responsible for a specific transcriptomic signature that may explain some of the features of APA carrying a *KCNJ5* mutations. This cellular model thus provides important new insight into the mechanisms leading to PA development, uncovering new mechanisms involved in the disease, thereby paving the way for new therapeutic approaches in primary aldosteronism.

The strengths of our study are multiple. First, the model provides a novel and highly valuable tool for studying the early events that occur following a mutational event leading to Na⁺ entry into the cells, such as mutations affecting *KCNJ5*[8] but also mutations affecting *SLC30A1*[15], *ATP1A1*[9,16] and *MCOLN3*[55]. Second, the cell model allows following dynamic changes in cell proliferation and function and to explore molecular mechanisms and specific transcriptional changes induced by mutations in PA. Importantly, these events cannot be studied in human tissues from patients with APA, as they only provide a snapshot at a given moment of the development of the disease. Indeed, we were able to show that while modification of steroid biosynthesis occurs very rapidly, alterations in the cellular sodium balance result in reduced cell proliferation and increased apoptosis, in accordance with data from the literature[18]. However, we cannot exclude the possibility that, at a later stage of development of APA, the constitutive activation of $Ca^{2+}$ signaling might stimulate cell proliferation. Another significant finding of our study was the identification of specific pathways and signatures associated with the modification of $Na^+$ influx. While some pathways, such as Aldosterone biosynthesis and secretion, calcium signaling and cell proliferation, were further explored, others remain to be investigated and will be the scope of future research. Moreover, this innovative cellular model offers a valuable tool for evaluating the molecular effects of novel therapeutic strategies for PA, particularly the application of novel aldosterone synthase inhibitors.

Limitations of our study are related to possible biases in interpreting proliferation studies in our model. Indeed, these cells are generated from H295R-S2 cells, which derive from an adrenocortical adenoma carrying TP53 and β-catenin mutations. Since β-catenin signaling is known to promote cell proliferation in various tumors, this model may be suboptimal for investigating positive effects on proliferation. Nevertheless, we have demonstrated that the activation of $Ca^{2+}$ signaling, through modulation of $Na^+$ entry into the cells is achievable within this context. Therefore, the presence of β-catenin mutations does not pose an obstacle to these specific investigations. An additional limitation may be the lack of the microenvironment of steroidogenic cells; further studies in integrated mouse models will be necessary to establish the natural history of PA development.

**Perspectives**

Using a chemogenetic approach allowing the modulation of sodium influx "on demand", we have generated a new cell model which replicates the main characteristics of *KCNJ5*-mutated APAs and provides new insights into the mechanisms responsible for their development. Our results show that modulation of intracellular ionic balance leads to cell membrane depolarization, activation of $Ca^{2+}$ signaling and enhanced steroid biosynthesis, associated with a decrease in cell viability and an increase in apoptosis, indicating that additional events may be required for the development of an APA with *KCNJ5* mutation.

This innovative model could be used to test the molecular impact of new treatments for PA *in vitro*. Additionally, applying this chemogenetic approach could also enable the development of a new inducible mouse model of PA, providing a valuable tool for dissecting the mechanisms underlying APA development and for evaluating new therapeutic strategies.

**Acknowledgments**

**Acknowledgments**

None

**Sources of Funding**

This work was funded through institutional support from INSERM, by the Agence Nationale pour la Recherche (ANR-18-CE93-0003-01), the Fondation pour la Recherche Médicale (EQU201903007864), the European Union's Horizon 2020 research and innovation programme under the Marie Sklodowska-Curie grant agreement No. 954798 (MINDSHIFT ITN), for which M-C.Zennaro. is principal investigator, and by a grant from the Leducq Foundation (18CVD05) for J-S.Hulot. Research in D. Penton's laboratory was financed by the Swiss National Foundation for Science (CRSII-222773) and by the University Research Priority Program of the University of Zurich (URPP) ITINERARE-Innovative Therapies in Rare Disesases.

**Disclosures**

None

**Novelty and relevance**

**What is new?**

Herein, we have developed a novel cell model in which intracellular $Ca^{2+}$ concentration can be modulated on demand, through modulation of sodium entry, reproducing the major characteristics of cells harboring a *KCNJ5* mutations.

We demonstrate that $Na^+$ entry into the cells results, on one hand, in cell membrane depolarization, $Ca^{2+}$ influx, activation of $Ca^{2+}$ signaling leading to increased expression of *CYP11B2* and stimulation of mineralocorticoid biosynthesis and, on the other hand, to a decrease in cell proliferation and an increase in apoptosis,

We show that $Na^+$ influx leads to a specific transcriptome signature that may explain some of the peculiar features of aldosterone producing adenoma carrying KCNJ5 mutations compared to other genetic abnormalities.

**What is relevant?**

While the role of *KCNJ5* mutations in promoting autonomous aldosterone biosynthesis has been clearly established, their role in promoting also abnormal cell proliferation remains to be established. Our study demonstrates that, *KCNJ5* mutations are not able to promote both autonomous aldosterone production and cell proliferation, suggesting that additional events may be required for adenoma development.

**Clinical/pathophysiological implications?**

This cellular model offers valuable new insights into the mechanisms leading to primary aldosteronism development, revealing novel processes involved in the disease and paving the way for new therapeutic approaches in PA.

**Supplemental material**

Expanded Materials and Methods

Tables S1-S15

Figures S1-S13

Table 1. Steroid profiles in H295R_S2 cells expressing the α7-5HT3-R in response to AngII ($10^{-8}$M), K$^+$ (12mM) and uPSEM-817 ($10^{-9}$ to $10^{-5}$M).

| Characteristics | Time | AngII ($10^{-8}$M) | K+ (12mM) | PSEM ($10^{-9}$M) | PSEM ($10^{-8}$M) | PSEM ($10^{-7}$M) | PSEM ($10^{-6}$M) | PSEM ($10^{-5}$M) |
|---|---|---|---|---|---|---|---|---|
| Pregnenolone | 8h | 2.622±0.176**** | 2.458±0.203**** | 1.071±0.025 | 1.035±0.024 | 1.269±0.031**** | 1.254±0.037*** | 1.489±0.066**** |
| | 24h | 0.441±0.023**** | 0.813±0.038 | 1.031±0.036 | 1.045±0.041 | 1.485±0.206 | 1.419±0.157* | 1.643±0.065*** |
| Progesterone | 8h | 4.005±0.366**** | 3.112±0.296** | 1.167±0.020 | 1.150±0.016 | 1.422±0.065*** | 1.453±0.065**** | 1.652±0.077**** |
| | 24h | 0.330±0.022**** | 0.533±0.036**** | 1.058±0.058 | 1.002±0.056 | 1.165±0.045 | 1.115±0.052 | 1.196±0.048* |
| 11-deoxycorticosterone | 8h | 4.100±0.478*** | 3.811±0.350** | 1.025±0.046 | 0.981±0.060 | 1.272±0.049* | 1.241±0.039* | 1.378±0.060** |
| | 24h | 0.409±0.044**** | 0.828±0.051 | 1.055±0.040 | 1.012±0.024 | 1.219±0.041** | 1.197±0.056** | 1.349±0.048**** |
| 18-hydroxy 11-deoxycorticosterone | 8h | 2.761±0.205*** | 2.585±0.238** | 1.122±0.019 | 1.114±0.023 | 1.260±0.055*** | 1.353±0.064**** | 1.455±0,038**** |
| | 24h | 2.467±0.193**** | 2.235±0.198**** | 0.982±0.038 | 0.973±0.034 | 1.234±0.041* | 1.271±0.081 | 1.404±0,101** |
| Corticosterone | 8h | 2.515±0.179**** | 2.424±0.245**** | 1.108±0.035 | 1.148±0.036 | 1.294±0.074 | 1.090±0.187 | 1.021±0.188 |
| | 24h | 2.329±0.268**** | 1.983±0.183** | 0.984±0.051 | 1.002±0.055 | 1.169±0.033* | 1.174±0.032** | 1.302±0.026**** |
| 18-hydroxycortisol | 8h | 0.794±0.090 | 0.921±0.112 | 0.944±0.057 | 0.754±0.126 | 0.910±0.053 | 0.894±0.037 | 1,314±0.249 |
| | 24h | 2.200±0.575 | 8.049±2.958** | 2.203±0.557 | 2.241±0.595 | 3.047±1.206 | 2.126±0.691 | 2.086±0.622 |
| 18-oxocortisol | 8h | 1.046±0.094 | 1.131±0.126 | 0.937±0.043 | 0.798±0.075 | 0.929±0.028 | 1.068±0.102 | 1.633±0.355 |
| | 24h | 2.200±0.575 | 8.049±2.958** | 2.203±0.557 | 2.241±0.595 | 3.047±1.206 | 2.126±0.691 | 2.086±0.622 |
| 18-hydroxycorticosterone | 8h | 1.927±0.172*** | 2.088±0.156**** | 1.125±0.033 | 1.125±0.052 | 1.231±0.051* | 1.003±0.108 | 0.835±0.063 |
| | 24h | 3.496±0.400**** | 2.456±0.341** | 0.989±0.074 | 1.004±0.094 | 1.177±0.157 | 0.891±0.102 | 0.818±0.085 |
| Aldosterone | 8h | 2.454±0.238** | 2.580±0.273*** | 0.914±0.040 | 0.927±0,038 | 1.024±0.054 | 1.187±0.063 | 1.388±0.062 |
| | 24h | 12.13±4.007**** | 6.046±1.657** | 0.946±0.043 | 1.035±0.082 | 1.495±0.211 | 1.459±0.138** | 1.436±0.135** |
| 17-hydroxyprogesterone | 8h | 2.761±0.351**** | 2.535±0.242*** | 1.187±0.027* | 1.141±0.035 | 1.334±0.059**** | 1.325±0.051**** | 1.589±0.041**** |
| | 24h | 0.243±0.009**** | 0.896±0.034** | 1.052±0.038 | 0.975±0.041 | 1.068±0.034 | 1.049±0.032 | 1.230±0.030** |
| 11-deoxycortisol | 8h | 1.586±0.147*** | 1.706±0.105*** | 1.026±0.033 | 0.995±0.056 | 1.087±0.038 | 1.064±0.035 | 1.079±0.039 |
| | 24h | 1.065±0.046 | 0.626±0.064**** | 1.016±0.025 | 0.985±0.030 | 1.096±0.041 | 1.073±0.052 | 1.060±0.046 |
| Cortisol | 8h | 1.093±0.061 | 1.066±0.056 | 1.119±0.082 | 1.157±0.058 | 1.126±0.081 | 1.051±0.057 | 1.003±0.055 |
| | 24h | 2.541±0.310**** | 1.923±0.251** | 0.974±0.033 | 1.034±0.041 | 1.161±0.080 | 1.194±0.095 | 1.199±0.076 |
| 21-deoxycortisol | 8h | 1.546±0.135* | 1.443±0.139 | 1.046±0.074 | 1.059±0.146 | 0.968±0.143 | 0.926±0.163 | 1.863±0.548 |
| | 24h | 0.852±0.121 | 1.954±0.115**** | 1.151±0.120 | 1.107±0.098 | 1.345±0.162 | 1.195±0.073 | 0.941±0.122 |
| Delta-4-androstenedione | 8h | 1.450±0.123* | 1.541±0.131** | 1.001±0.033 | 0.978±0.085 | 1.113±0.039 | 1.069±0.034 | 1.246±0.038** |
| | 24h | 0.482±0.046**** | 1.146±0.055 | 1.049±0.032 | 1.003±0.027 | 1.098±0.047 | 0.985±0.039 | 1.104±0.041 |

Results are expressed as fold induction over untreated cells expressing the $\alpha$7-5HT3-R (set as 1, not shown) and represent mean±SEM of the three clones, compared with ANOVA followed by Bonferroni when data presented for Gaussian distribution and Kruskal–Wallis followed by Dunn's test for non-Gaussian distribution. Cells treated with AngII and $K^+$ and cells treated with uPSEM-817 (PSEM) were analyzed separately. n=9/condition except for 18-oxocortisol (n=6) and 18-hydroxycortisol (n=6).
****, $p \leq 0.0001$; ***, $p \leq 0.001$; **, $p \leq 0.01$, *, $p \leq 0.05$

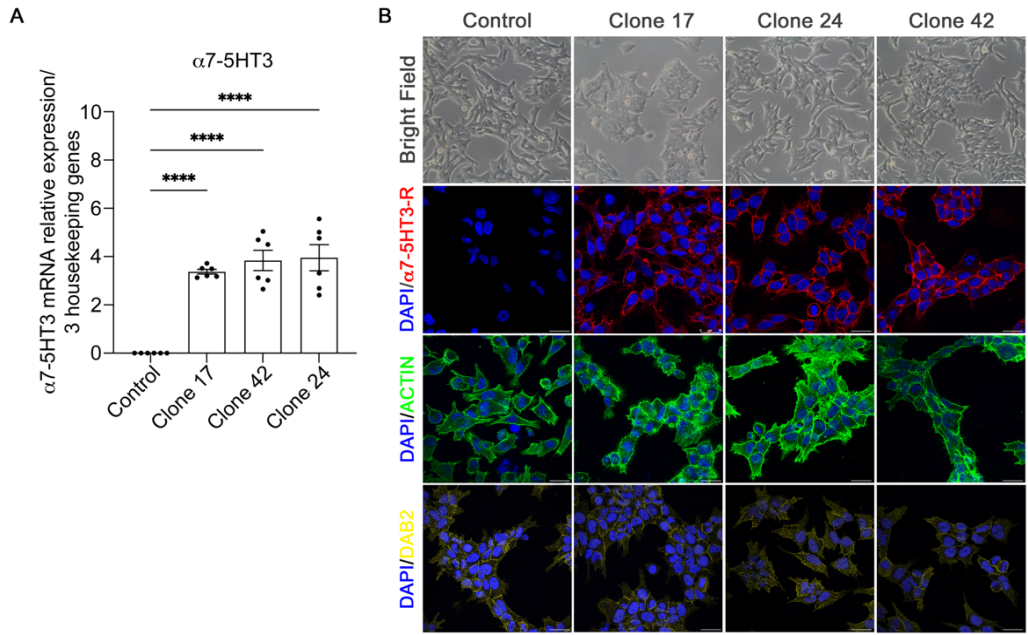

**Figure 1. Morphological characterization of α7-5HT3-R expressing cells.** The expression of α7-5HT3-R was investigated in control cells, expressing an empty vector, and in three selected clones, 17, 24 and 42. **(A)** mRNA expression of α7-5HT3-R was investigated by RT-qPCR, n=6 for each clone **(B)** Morphological characteristics of control and α7-5HT3-R expressing cells were evaluated by immunofluorescence. Bright field images revealed no structural difference between cells. DAPI nuclear staining is shown in blue, α-bungarotoxin (α7-5HT3-R) in red, Phalloidin (actin) in green and dab2 in yellow. *P* values were determined by one-way ANOVA followed by Bonferroni post-hoc comparisons tests, ***, p<0,0001, Bar represents 20μm

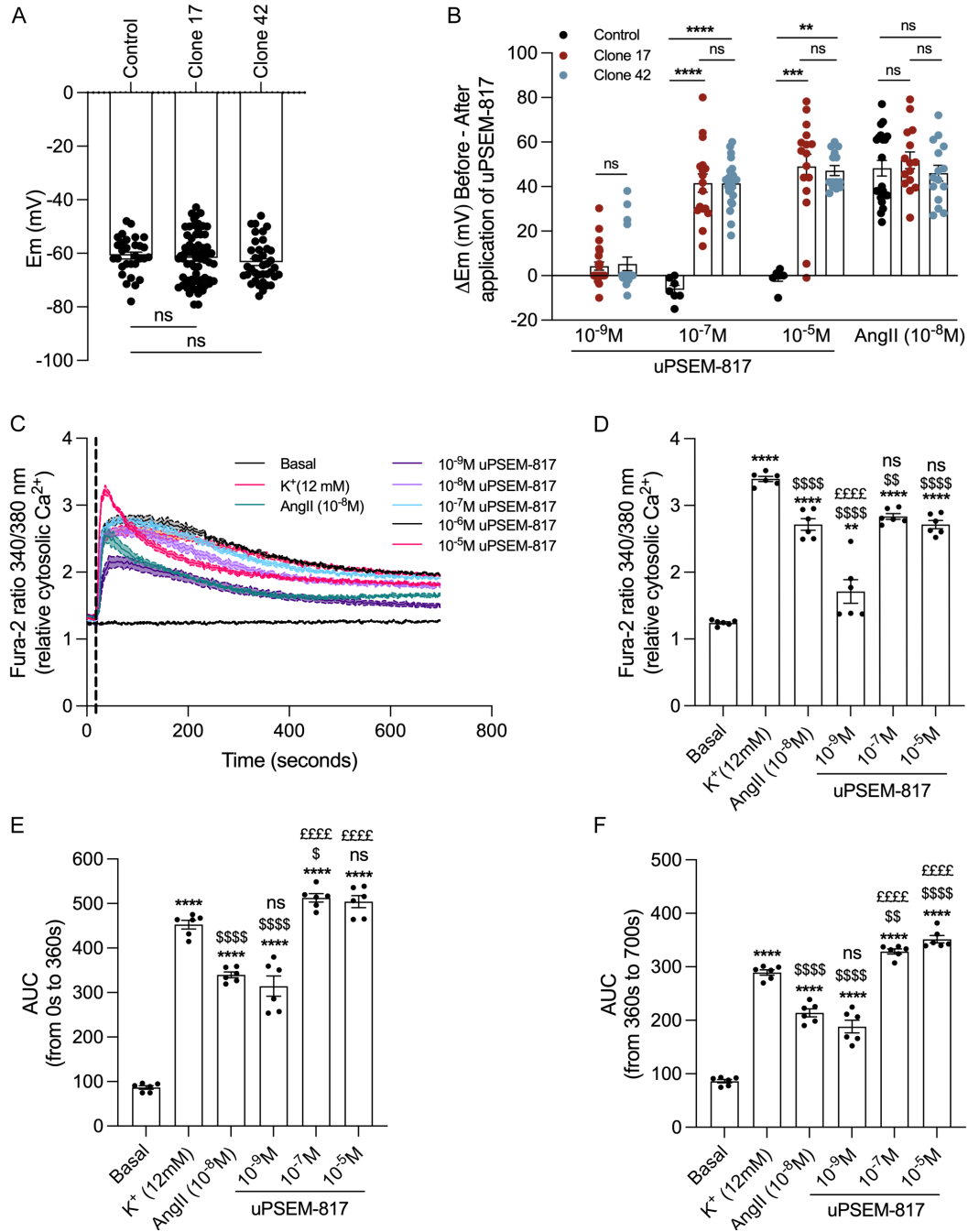

**Figure 2. Functional characterization of α7-5HT3-R expressing cells. (A and B)** Perforated patch clamp experiments were performed on clones 17 and 42 to determine the membrane potential of cells in basal condition, n=15-21 **(A)** or after stimulation with different concentrations of uPSEM-817 ($10^{-9}$, $10^{-7}$ and $10^{-5}$M) and AngII ($10^{-8}$M), n=7-23 **(B)**. **(C)** $Ca^{2+}$ entry into the cells was evaluated using Fura-2 AM assay for 700s. Representative traces of intracellular $Ca^{2+}$ responses to $10^{-8}$M AngII, 12mM $K^+$ and $10^{-9}$ to $10^{-5}$M uPSEM-817 (Clone 17), n=6 **(D)** Determination of the maximum Fura-2 ratio 340/380nm in response to $10^{-8}$M AngII, 12mM $K^+$ and $10^{-9}$, $10^{-7}$ and $10^{-5}$M uPSEM-817 (Clone 17), n=6. **(E)** Area under the curve (AUC) was determined to assess whether treatment with PSEM induced $Ca^{2+}$ entry into the cells. The AUC was determined between 0 and 360s to determine the intracellular $Ca^{2+}$ variation during the peak response (left panel) and between 360s and 700s to determine the steady-state response (right panel) (Clone 17). n=6, *P* values were determined by t-test or one-

way ANOVA followed by Bonferroni post-hoc comparisons tests, * basal vs $K^+$, AngII or PSEM, $ $K^+$ vs AngII or PSEM, £ AngII vs PSEM $; p<0,05; $$, p<0,01, ns AngII vs $10^{-9}$M PSEM (left and right panel), $K^+$ vs $10^{-5}$M PSEM (right panel); ****, p<0,0001; ns: not significant. The dotted line indicates the time of injection of uPSEM-817, AngII or $K^+$.

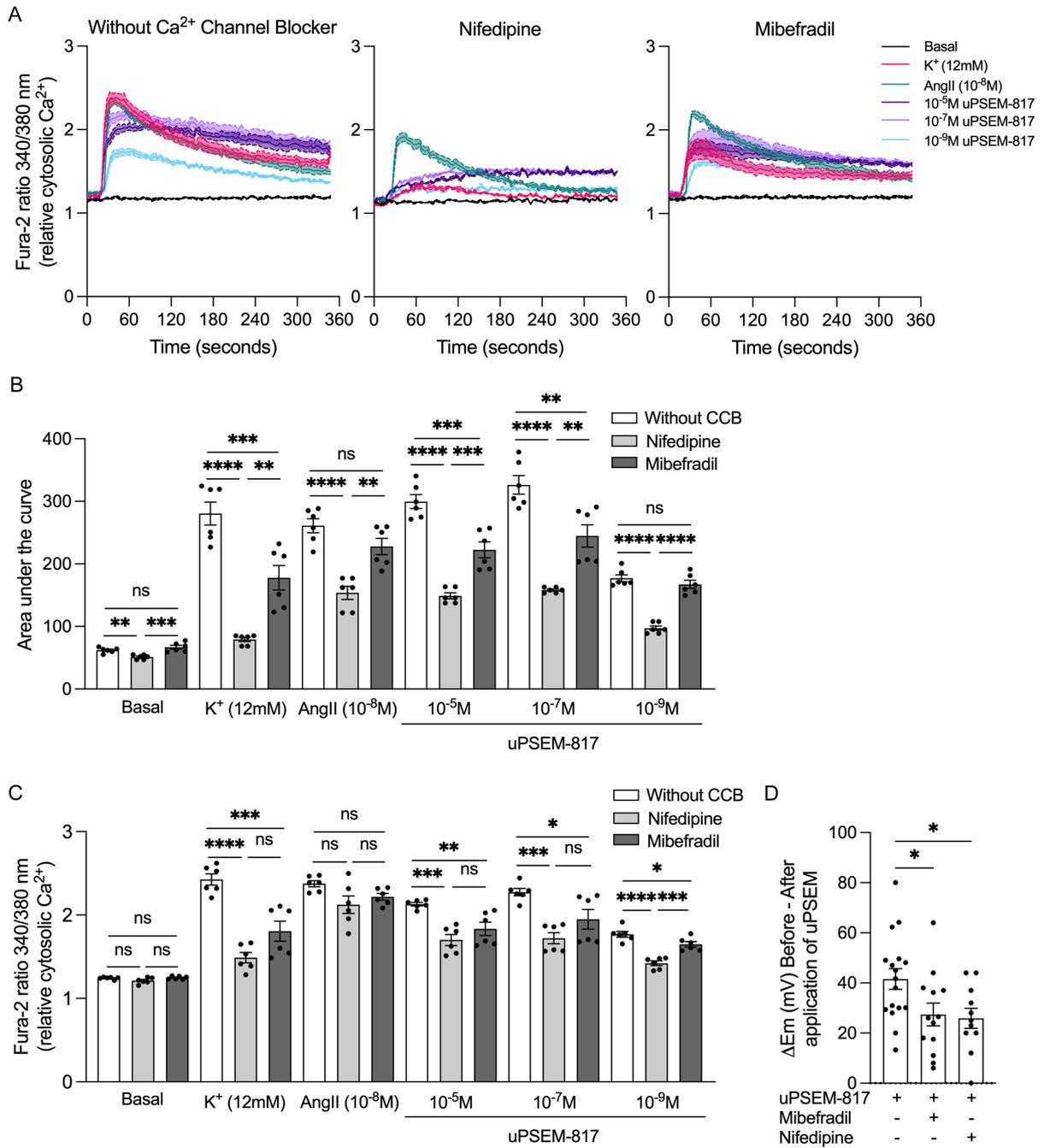

**Figure 3. Inhibition of $Ca^{2+}$ entry using Nifedipine or Mibefradil. (A)** Representative traces of intracellular $Ca^{2+}$ responses to $10^{-8}$M AngII, 12mM $K^+$ and $10^{-9}$, $10^{-7}$ and $10^{-5}$M uPSEM-817 after pre-treatment with Nifedipine (left), or Mibefradil (right) (Clone 17); n=6. **(B)** Determination of the maximum Fura-2 ratio 340/380nm in response to uPSEM-817, AngII and $K^+$ after pre-treatment with nifedipine (red) or mibefradil (blue) or without pre-treatment (black) (Clone 17); n=6. **(C)** Illustration of the $[Ca^{2+}]_i$ signaling, illustrated by the determination of the AUC assessed for 360s, in response to uPSEM-817, AngII and $K^+$ after pre-treatment with nifedipine (red) or mibefradil (blue) or without pre-treatment (black) (Clone 17); n=6. **(D)** Path clamp recordings of cells treated with uPSEM-817 and/or nifedipine or mibefradil (Clone 17). n=11-17, *P* values were determined by one-way ANOVA followed by Bonferroni post-hoc comparisons tests, *, p<0.05; **, p<0.01; ***, p<0.001; ****, p<0.0001

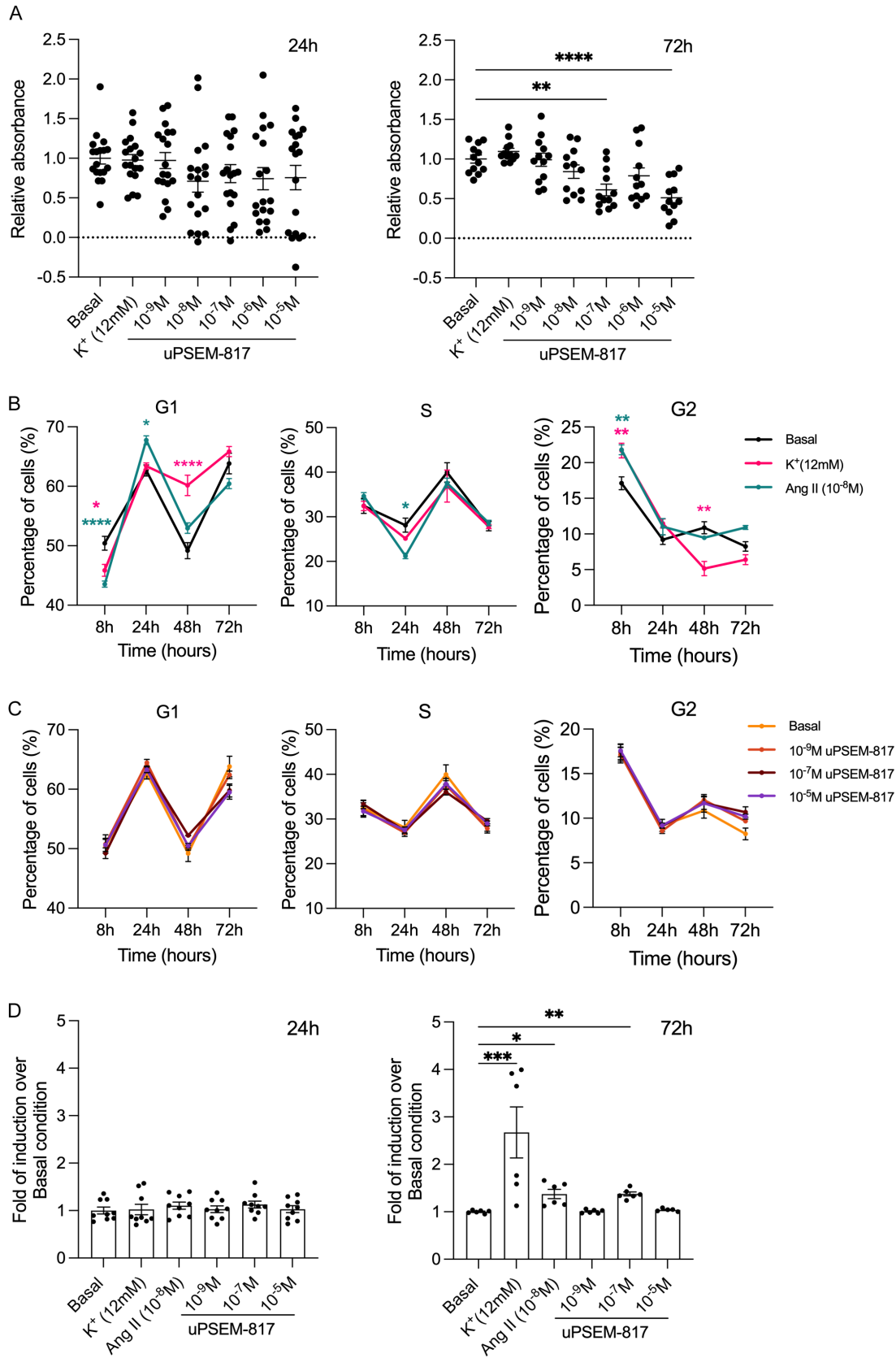

**Figure 4. Effect of uPSEM-817 on cell proliferation and apoptosis.** (**A**) Cell viability was measured on cells treated for 24h or 72h with 12mM K$^+$ or uPSEM-817 by MTS assay (Clone 17). (**B**) Cell cycle distribution measured by FACS using propidium iodide in response to 10$^{-8}$M AngII and 12mM K$^+$ (upper panel) and 10$^{-9}$, 10$^{-7}$ and 10$^{-5}$M uPSEM-817 (lower panel)

(Clone 17). (**C**) Apoptosis determined by the proportion of cells in sub-G1 phase (Clone 17). (**D-E**) Proportion of cells in the sub-G1 phase after 24h (**D**) or 72h (**E**) in response to 12mM $K^+$, $10^{-8}$M AngII and $10^{-9}$, $10^{-7}$ and $10^{-5}$M uPSEM-817. n=6 (cell cycle), n=12 (proliferation), *P* values were determined by one-way ANOVA followed by Bonferroni post-hoc comparisons tests, *, p<0.05; **, p<0.01; ****, p<0.0001

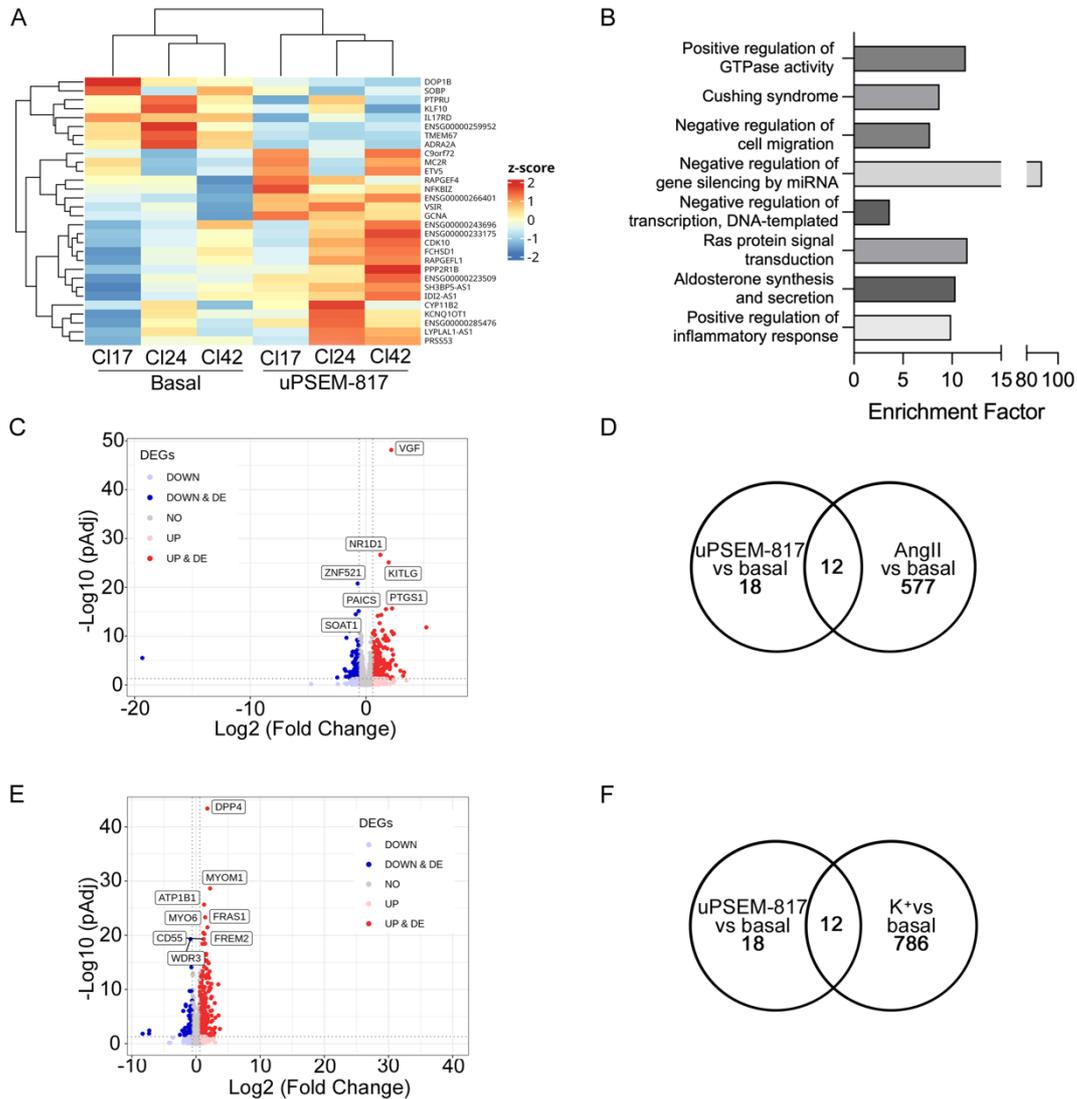

**Figure 5. Gene expression profiles of H295R-S2 α7-5HT3-R cells treated 24h with $10^{-7}$M uPSEM-817, $10^{-8}$M AngII or 12mM $K^+$.** (**A**) Hierarchical clustering of samples using the 28 differentially expressed genes in cells treated or not with $10^{-7}$M uPSEM-817. (**B**) Biological process enrichments determined using the list of differentially expressed genes in cells treated or not with $10^{-7}$M uPSEM-817. (**C**) Volcano plot showing the 4932 differentially in response to $10^{-8}$M AngII. Differential expressed genes are highlighted as blue (down-regulated) or red (up-regulated) dots. (**D**) Venn diagram representing the common and different genes differentially expressed in response to $10^{-8}$M AngII and $10^{-7}$M uPSEM-817. (**E**) Volcano plot showing the 728 differentially in response to 12mM $K^+$. The x-axis is the Log2 fold change between the two conditions; the adjusted p value based on $-\log_{10}$ is reported on the y-axis. Genes significantly different are highlighted as blue (down-regulated in cells treated with 12mM $K^+$) or red (up-regulated in cells treated with 12mM $K^+$) dots. (**F**) Venn diagram representing the common and different genes differentially expressed in response to 12mM $K^+$ and $10^{-7}$M uPSEM-817.